\documentstyle[aps,preprint]{revtex}
\begin{document}

\title{Analytically solvable mean-field potential\\
for stable and exotic nuclei}
\author{
M.V.~Stoitsov,$^{1}$ S.S.~Dimitrova,$^{1}$
S.~Pittel,$^{2}$ P.~Van Isacker,$^{3}$ and A.~Frank$^{4,5}$}
\address{
$^{1}$Institute of Nuclear Research and Nuclear Energy,
Bulgarian Academy of Sciences,
Sofia--1784, Bulgaria \\
$^{2}$Bartol Research Institute, University of Delaware,
Newark, DE 19716 USA \\
$^{3}$Grand Acc\'el\'erateur National d'Ions Lourds,
BP 5027, F-14021 Caen Cedex, France \\
$^4$Instituto de Ciencias Nucleares, U.N.A.M.,
A.P. 70-543, 04510 M\'{e}xico, D.F., M\'{e}xico \\
$^5$Instituto de F\'\i sica, Laboratorio de Cuernavaca, U.N.A.M.,
A.P. 139-B, Cuernavaca, Morelos, M\'{e}xico \\
}
\date{\today}
\maketitle

\begin{abstract}

Slater determinants built from the single-particle wave functions of the
analytically solvable Ginocchio potential are used to approximate the
self-consistent Hartree-Fock solutions for the ground states of nuclei. The
results indicate that the Ginocchio potential provides a good parametrization of
the nuclear mean field for a wide range of nuclei, including those at the limit
of particle stability. \end{abstract}

\begin{center}
{\bf PACS numbers:} 21.60.Jz, 21.60.Cs, 21.10.Dr, 21.10.Ft, 21.10.Gv
\end{center}

\draft

\newpage

Although the harmonic-oscillator potential is a crude approximation to the true
nuclear mean field, it is frequently used in applications of the shell model of
atomic nuclei because the radial dependence of its eigenfunctions is known in
closed form, allowing an understanding of many nuclear properties from a
simple---albeit approximate---perspective \cite{TAL93}. More realistic
approximations to the nuclear mean field are, of course, available and the most
prominent among them, the Woods-Saxon potential, provides a parametrization in
terms of a radial extent and a surface diffuseness, allowing a greater
flexibility than the single length parameter of the harmonic oscillator. This
greater flexibility, however, occurs at the cost of analyticity.

With the advent of experimental facilities that accelerate radioactive species
\cite{MUE93}, the need for a flexible parametrization of the nuclear mean field
becomes all the more pressing. Indeed, with so many more nuclei available for
experimental detection a wider variety in the mean-field potential is expected.
Much work is currently in progress to develop appropriate theoretical tools for
describing nuclei up to the line of particle stability. In light nuclei,
extensions of traditional shell-model techniques have confirmed the necessity to
modify the single-particle shell-model potential when dealing with nuclei far
from stability \cite{OTS93}. In heavier nuclei, the principal developments
so far
have been in the context of mean-field theory \cite{DOB84} and the use of
shell-model techniques \cite{BEN96} will likewise require renewed consideration
of an appropriate shell-model potential. Ideally, it should be chosen so as to
properly reflect the self-consistent single-particle potential felt by each
nucleon, including those modifications that are expected for very weakly bound
nucleons near the nuclear surface. Already, some work aimed at analyzing the
local equivalent of the full Hartree-Fock (HF) potential in nuclei with a large
neutron excess suggests that the potential develops an increasingly large
diffuseness as the neutron excess grows \cite{DOB96,FUK93}. Clearly, such
phenomena elude a shell-model description based on a harmonic-oscillator
potential.

Some time ago, an analytically solvable potential depending on several
parameters
was introduced by Ginocchio \cite{GIN85}. For certain combinations of these
parameters, this potential mimics the usual Woods-Saxon potential with a small
diffuseness as occurs in nuclei in the valley of stability. For other parameter
combinations, it admits very different shapes, in some cases reducing to the
highly diffuse P\"oschl-Teller potential familiar in molecular physics. Because
of this versatility, the Ginocchio potential may be a useful approximation to the
spherical single-particle shell-model potential of nuclei close to as well as far
from stability.

In this Letter, we study the possible usefulness of the Ginocchio potential in
nuclear structure physics. More specifically, we ask the question: Does a Slater
determinant built from the single-particle eigenstates of this potential---with
the parameters chosen variationally---reproduce the properties of a
self-consistent Hartree-Fock (SCHF) calculation? For the sake of terminology, we
refer to this variational procedure to generate the optimum potential as the
Ginocchio-Hartree-Fock (GHF) procedure; we also compare its results to a
variational procedure (called the harmonic-oscillator Hartree-Fock or HOHF)
defined similarly but with reference to harmonic-oscillator single-particle
states. This analysis is carried out using a Skyrme hamiltonian and for a
representative set of doubly-magic nuclei, including nuclei both within and
outside the traditional valley of stability. The comparisons include such key
nuclear characteristics as binding energies, density distributions, and  rms
radii.

The main features of the SCHF theory based on Skyrme-type effective forces are
well known \cite{VAU72,BRA85}. The total HF energy, taken as an expectation value
of the nuclear Hamiltonian $\hat{H}$ over a trial Slater determinant,
involves a sum of a Skyrme energy density ${\cal E}_{\rm Sky}$ and a Coulomb
energy density ${\cal E}_{\rm Coul}$. The Skyrme energy density is expressed for
spherical even-even nuclei in terms of local nucleon densities, kinetic energy
densities, and spin-orbit densities, all defined in terms of the variational
single-particle states used to build the Slater determinant. The Coulomb energy
density, which depends on the local proton density, contains both a direct and an
exchange term, the latter usually taken in Slater approximation.

To set the stage for a discussion of the GHF procedure, it is useful to make a
few brief remarks about the Ginocchio potential \cite{GIN85}. The potential
involves four parameters, $s$, $\lambda$, $a$, and $\nu$, which are associated to
different geometrical features.  Thus, $s$ is the `scaling parameter', $\lambda$
is the `shape parameter', $a$ is the `effective mass parameter' and $\nu$ is the
`depth parameter'.

In general, each `$lj$ shell' of the Ginocchio potential may have its own set of
parameters $s$, $\lambda$, $a$, and $\nu$, which furthermore may be different for
neutrons and protons. For any such choice of the parameters, the potential is
analytically solvable. Formulae for its eigenenergies and corresponding
eigenvectors are detailed in Ref.\ \cite{GIN85} and will not be repeated here.

For a variational procedure to be useful, it should not have too many parameters.
Thus, while the full parameter flexibility of the Ginocchio potential may be a
benefit in some regards, it is a hindrance for our purposes and a limitation in
the number of parameters is essential. We will present results of calculations
with two sets of variational parameters: (1) In the calculations denoted G4, we
assume the four parameters, $s$, $\lambda$, $a$, and $\nu$, to be entirely
global, with the same values for all neutron and proton single-particle states;
(2) More detailed calculations (denoted G8) are for an eight-parameter potential
in which we allow distinct parameters for neutrons and protons.

With this as background, the GHF procedure involves (a) evaluating the expectation
value of the (Skyrme-type) nuclear Hamiltonian over a Slater determinant built
from the Ginocchio single-particle eigenstates as a function of the parameters of
the potential (four for G4 and eight for G8), and then (b) minimizing
this energy functional with respect to its parameters.

The HOHF procedure is formulated in the same way, the only difference being that
the trial Slater determinants are constructed from harmonic-oscillator
single-particle wave functions. The energy functional is then minimized with
respect to two parameters, the harmonic-oscillator length parameters for neutrons
$\alpha_n$ and for protons $\alpha_p$.

We have carried out HOHF, GHF, and SCHF calculations for a sequence of
doubly-magic nuclei. This includes several within the traditional valley of
stability -- $^{16}$O, $^{40}$Ca, $^{56}$Ni, and $^{208}$Pb -- and one exotic
nucleus $^{78}$Ni. The HOHF analysis, however, was limited to the nuclei
$^{16}$O, $^{40}$Ca and $^{208}$Pb. All calculations were based on the original
Skyrme parametrization (SI)\cite{VAU72}. Finally, as in the earliest calculations
based on the SI force \cite{VAU72}, we omitted the exchange contribution to
${\cal E}_{\rm Coul}$.

Our results for ground-state energies are given in Table~I.  A few observations
are in order. (1) The GHF ground-state energies are lower than those of HOHF,
indicating that the Ginocchio wave functions are better in the variational sense
than the harmonic-oscillator wave functions. This is not very surprising since
the Ginocchio potential (even in the case G4) contains more parameters. The same
remark applies when comparing the different calculations based on the Ginocchio
potential, namely that the ground-state energies are lower in G8 than in G4. As
expected, the SCHF ground-state energies are lower than those in GHF and HOHF
because of the variational nature of the calculations. (2) For $N=Z$ nuclei,
there is no significant difference between the G4, and G8 energies. In nuclei with
a substantial neutron excess, however, the G8 calculation is clearly superior.
This is also evident from the parameters found in G8 (see Table~II): while for
$N=Z$ nuclei the corresponding neutron and proton parameters are nearly equal,
there are considerable differences when $N>Z$. (3) In all the cases considered,
the G8 ground-state energies are within 2\% of the SCHF results.

Table~III gives the corresponding results for neutron and proton rms radii.
Similar remarks as for ground-state energies apply. (1) The GHF results are in
very good agreement with those of the SCHF procedure, both for light and heavy
nuclei. Small deviations, not larger than 0.35\%, are seen for the neutron rms
radius only. This is in stark contrast to HOHF, where (as expected)
significant deviations from SCHF occur. (2) For $N=Z$ nuclei, the
calculation G4 suffices; for nuclei with a sizable neutron excess, significant
improvements are obtained by allowing different parameters for neutrons and
protons (G8).

Fig.~1 shows the neutron and proton local density distributions for several of
the stable doubly-magic nuclei we have considered. The GHF results were obtained
with the G8 parametrization of the Ginocchio potential. We make the following
observations: (1) The GHF and SCHF densities agree almost perfectly in the
surface region. The main reason is that the Ginocchio wave functions  have an
appropriate exponential fall off, as shown in \cite{GIN85}. In contrast, as
is also evident from Fig.~1, the Gaussian asymptotic behavior of
harmonic-oscillator wave functions leads to significant deviations in the HOHF
surface density from that in SCHF. This observation also explains the nice GHF
results for binding energies and rms radii. Since the densities and energy
densities are weighted by $r^2 dr$, the main contribution to these nuclear
characteristics comes from the nuclear surface, where GHF is evidently superior
to HOHF. {\it The fact that the GHF procedure reproduces nuclear properties in the
surface region so accurately suggests that the single-particle wave functions so
obtained should be especially suitable in shell-model studies.} (2) The HOHF and
GHF densities differ significantly from the SCHF results at small $r$, namely in
the region of density shell oscillations. Both HOHF and GHF exhibit oscillations
at small $r$ that are too large -- the former more so than the latter --
especially for heavier nuclei. An improvement in the GHF local densities at small
$r$ might perhaps be achieved by permitting different $\nu$ parameters
for single-particle states with $j=l+\frac{1}{2}$ and $j=l-\frac{1}{2}$.

To study the effect of a neutron excess, we compare in Fig.~2 the neutron and
proton local density distributions in $^{56}$Ni and $^{78}$Ni as obtained in the
GHF calculation G8. The Ginocchio potential is able to accommodate nuclei with
very different densities, ranging from those in which the neutron and proton
densities are nearly identical to others where the neutrons develop a `skin'. The
corresponding $s$-wave potentials for neutrons and protons are shown in Fig.~3.
Note the difference in shape of the two potentials in $^{78}$Ni: the neutron
potential has `softer' edges, resembling a P\"oschl-Teller (or perhaps a
highly-diffuse Woods-Saxon) potential.

In summary, we have studied the possible usefulness of the Ginocchio potential as
an effective local potential for the shell model by comparing the results of
variational calculations based on Slater determinants built from such wave
functions with those of self-consistent HF. The results are very encouraging.
They suggest that the GHF method is able to reproduce remarkably well the full HF
results, especially in the vicinity of the nuclear surface, all the while keeping
a relatively simple analytic form for the potential. Most importantly, the
Ginocchio potential seems be equally suitable for traditional stable nuclei and
for relatively weakly-bound nuclei far from stability. This opens up the
possibility of meaningfully extending the shell-model methodology to exotic
nuclei.

In the future, we plan to look into the possible use of more realistic effective
hamiltonians, including both improved zero-range interactions of the Skyrme type
\cite{DOB84} as well as finite-range interactions \cite{DEC80}. When dealing with
finite-range interactions, the self-consistent HF potential is difficult to cast
in local form, so that our procedure may turn out to be particularly useful
there. In that context, the connection of our approach to recent work of
relevance to atomic clusters \cite{BUL95} should be noted. Eventually, we would
like to apply these methods to (spherical) open-shell nuclei as well, where it
will be necessary to include BCS pairing correlations. We should, of course,
include the exchange contribution to the Coulomb energy in fully realistic
applications. Finally, we also plan to investigate the relation between our
proposed GHF method and the Extended Thomas Fermi approximation \cite{TON84}.

This work was supported in part by the National Science Foundation under Grant
Nos.~PHY-9600445 and INT-9415876, by the Bulgarian National Foundation for
Scientific Research under contract No.~$\Phi$-527, by the French Centre National
de Recherche Scientifique, by DGAPA-UNAM under project IN105194 and by the
European Union under contract No.~CI1*-CT94-0072. The authors wish to acknowledge
fruitful discussions with W.~Nazarewicz, D.~Dean, W.~Satula, M.~Ploszajczak and
J.~Dukelsky.


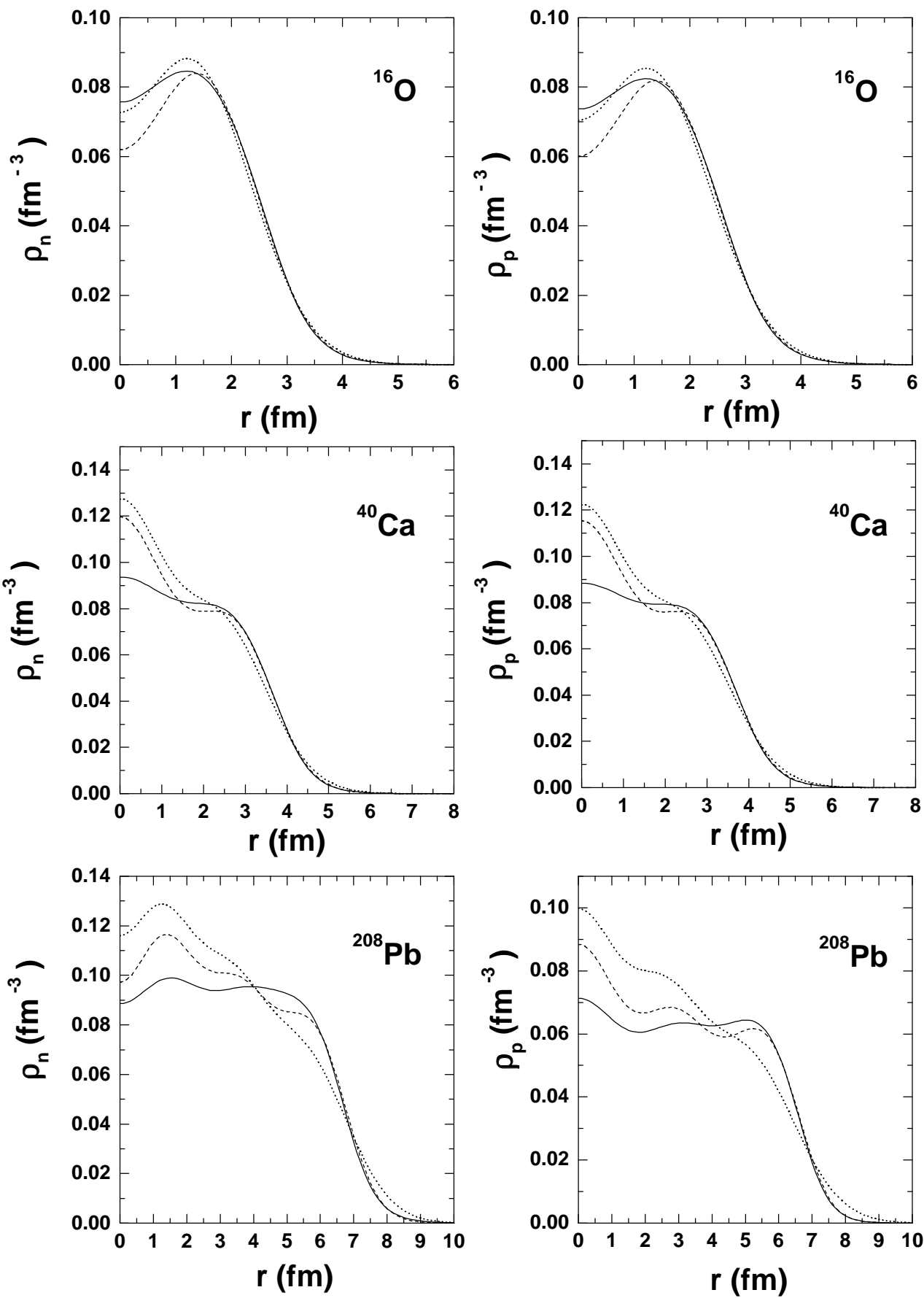
\begin{figure}
\caption[]{Neutron $\rho_n(r)$ and proton $\rho_p(r)$
local density distributions
as obtained in HOHF (dotted), GHF/G8 (dashed), and SCHF (solid) calculations.}
\end{figure}

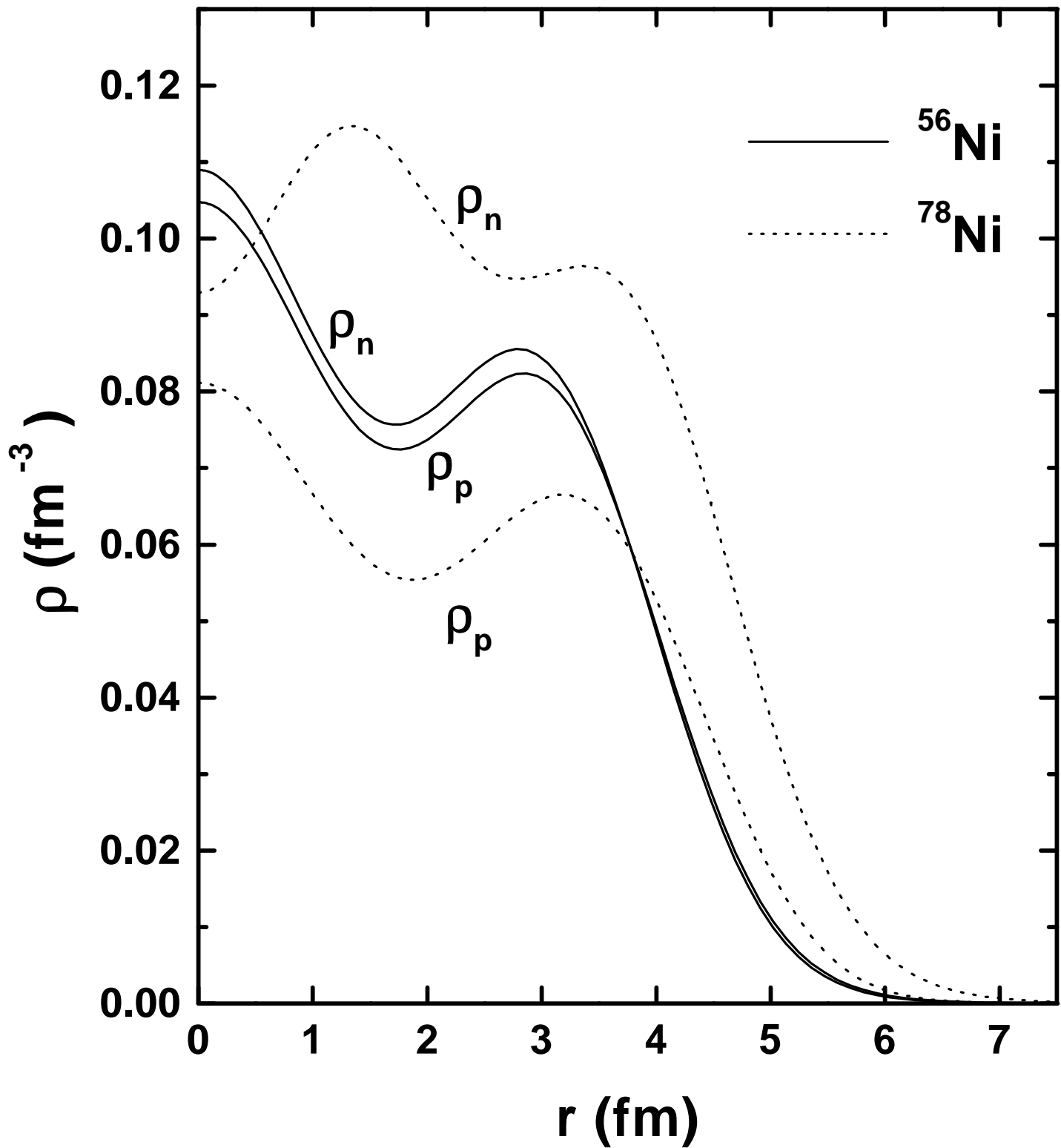
\begin{figure}
\caption[]{Neutron $\rho_n(r)$ and proton $\rho_p(r)$
local density distributions in $^{56}$Ni (solid) and $^{78}$Ni (dotted)
as obtained in the GHF calculation G8.}
\end{figure}

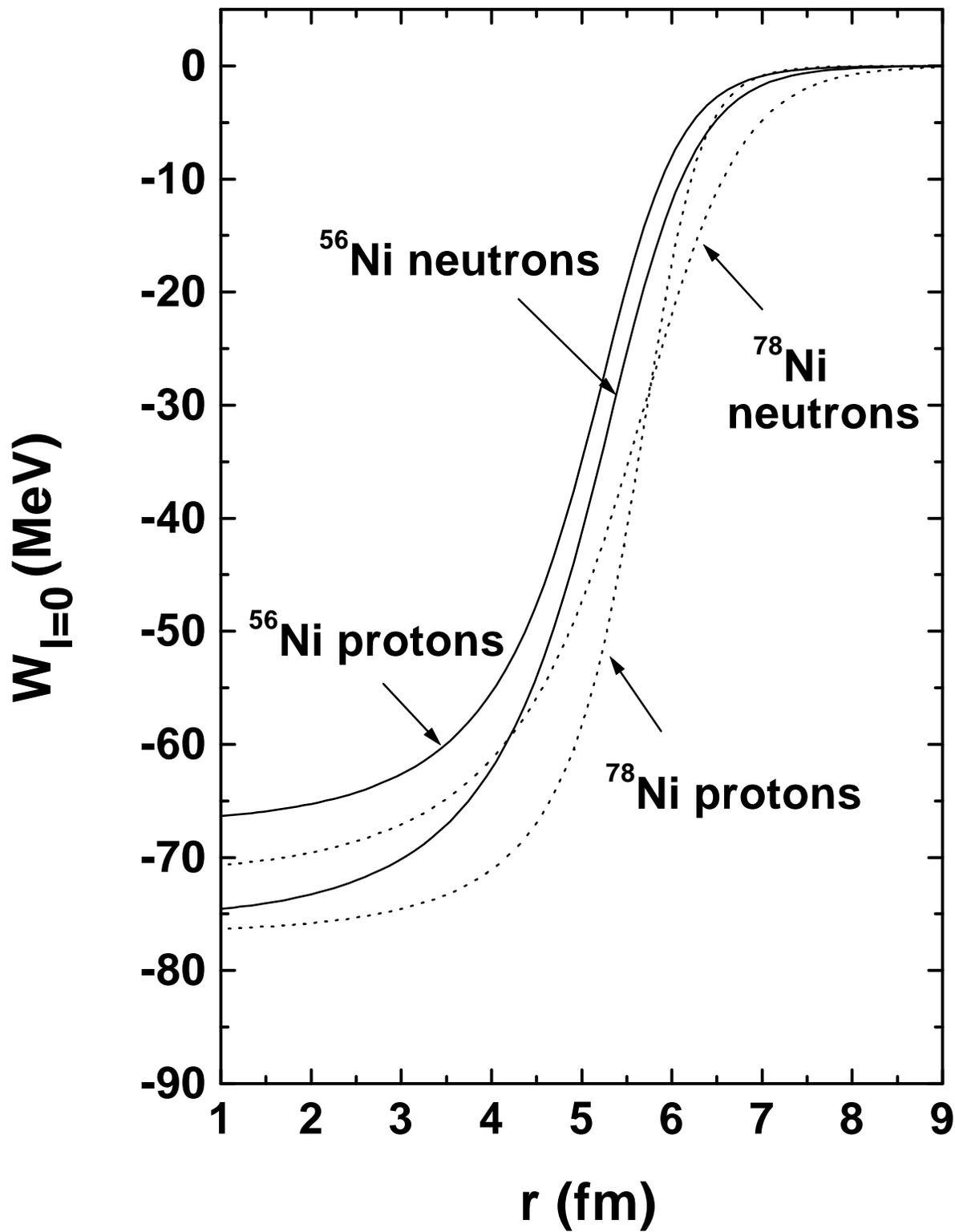
\begin{figure}
\caption[]{The $s$-wave potential for neutrons and protons
in $^{56}$Ni (solid) and $^{78}$Ni (dotted)
corresponding to the parameters of the GHF calculation G8
given in Table~III.}
\end{figure}

\begin{table}
\caption{Results for ground-state energies $E$ (in MeV)}
\begin{tabular}{lcccc}
&\multicolumn{4}{c}{$E$} \cr
\cline{2-5}
&HOHF&G4&G8&SCHF \cr
\cline{1-5}
$^{16}$O     & $-128.2$   & $-129.9$  & $-130.0$   & $-131.5$
\cr
$^{40}$Ca    & $-331.5$   & $-341.3$  & $-341.6$   & $-345.6$
\cr
$^{56}$Ni    & $-$        & $-489.8$  & $-489.8$   & $-498.7$
\cr
$^{78}$Ni    & $-$        & $-646.4$  & $-649.0$   & $-660.7$
\cr
$^{208}$Pb   & $-1469.3$  & $-1598.6$ & $-1609.2$  &$-1641.1$
\cr
\end{tabular}
\end{table}

\begin{table}
\caption{Variational parameters
as obtained in the GHF calculation G8$^{\rm a}$}
\begin{tabular}{lddddd}
           &$^{16}$O&$^{40}$Ca&$^{56}$Ni&$^{78}$Ni&$^{208}$Pb\\
\tableline
$s_n$      & 0.106  & 0.067   & 0.069   & 0.066   & 0.033    \\
$s_p$      & 0.105  & 0.064   & 0.066   & 0.042   & 0.022    \\
$\lambda_n$& 3.678  & 4.627   & 3.917   & 3.784   & 5.740    \\
$\lambda_p$& 3.714  & 4.878   & 4.138   & 6.097   & 9.033    \\
$a_n$      & 0.193  & 0.138   & 0.004   & 0.000   & 0.000    \\
$a_p$      & 0.151  & 0.131   & 0.000   & 0.124   & 0.001    \\
$\nu_n$    & 3.452  & 4.878   & 6.545   & 6.895   & 9.488    \\
$\nu_p$    & 3.400  & 4.447   & 6.029   & 6.504   & 8.434    \\
\end{tabular}
\tablenotetext[1]{The parameters $s_n$, $s_p$, $\alpha_n$, and $\alpha_p$
are in fm$^{-1}$; other parameters are dimensionless.}
\end{table}

\begin{table}
\caption{Results for the neutron and proton rms radii $R_n$ and $R_p$ (in
fm)}
\begin{tabular}{lccccccccc}
&\multicolumn{4}{c}{$R_n$}&
&\multicolumn{4}{c}{$R_p$} \cr
\cline{2-5} \cline{7-10}
&HOHF&G4&G8&SCHF &&HOHF&G4&G8&SCHF \cr
\cline{1-10}
$^{16}$O     &    2.56 &     2.55 &     2.54 &     2.53
            &&    2.58 &     2.55 &     2.56 &     2.56 \cr
$^{40}$Ca    &    3.32 &     3.29 &     3.27 &     3.27
            &&    3.36 &     3.29 &     3.31 &     3.31 \cr
$^{56}$Ni    &    ---  &     3.59 &     3.57 &     3.56
            &&    ---  &     3.59 &     3.62 &     3.61 \cr
$^{78}$Ni    &    ---  &     4.08 &     4.05 &     4.07
            &&    ---  &     3.82 &     3.86 &     3.85 \cr
$^{208}$Pb   &    5.58 &     5.51 &     5.47 &     5.49
            &&    5.47 &     5.32 &     5.38 &     5.38 \cr
\end{tabular}
\end{table}

\end{document}